\documentclass{ptapap}
\def \ptaJournalDate{to be published}
\def \ptaJournalVolume{vol. 2}
\def \ptaArticleURL{}
\def \ptaJournalName{PTA Proceedings}
\def \ArticleFirstPage{1}
\author{Marcin Semczuk}[CAMK,UW]
\author{Ewa L. {\L}okas}[CAMK]
\affil[CAMK]{Nicolaus Copernicus Astronomical Center, Bartycka 18, 00-716 Warszawa, Poland}
\affil[UW]{Warsaw University Astronomical Observatory, Al. Ujazdowskie 4, 00-478 Warszawa, Poland}

\title{Simulations of tidally induced spiral arms}

\begin{document}

\maketitle

\begin{abstract}
The origin of grand design spiral structure in galaxies is still under debate but one of
promising scenarios involves tidal interactions. We use $N$-body simulations to study the evolution
of a Milky Way-size galaxy in a Virgo-like cluster. The galaxy is placed on a typical eccentric
orbit and evolved for 10 Gyr. We find that grand design spiral arms are triggered by pericenter
passages and later on they wind up and dissipate. The arms formed in the simulations are
approximately logarithmic, but are also dynamic, transient and recurrent.
\end{abstract}

\section{Introduction}

Formation and evolution of spiral arms in galaxies remains one of the great unsolved
problems in modern astrophysics. Several theories aim to explain this phenomenon, but none of them
is generally believed to be complete and universally applicable. One of the scenarios proposes that
spiral arms originate from tidal interactions with another body, e.g. a galaxy of similar size.
This idea was first explored in the seminal paper by \cite{holmberg}.

More recent studies using full $N$-body simulations \citep[e.g.][]{oh15}
support the hypothesis that tidal encounters induce grand design, two-armed spiral
structure (as seen e.g. in M51). \citet{oh15} discussed the results of
$N$-body simulations of the satellite triggering the formation of spiral arms in a larger, disky galaxy. They
found the arms to be approximately logarithmic in shape and decaying
with time.

In this paper we present first results of $N$-body simulations of a Milky Way-size galaxy orbiting
in the Virgo-like cluster. We find that the formation of grand design spiral arms
in a galaxy can be triggered by tidal interactions with the cluster-size dark
matter halo.

\section{The simulations}

In our simulations the galaxy was modelled as an exponential stellar
disk embedded in an NFW (\citealt{nfw}) dark matter halo.
The model had properties similar to the Milky Way model MWb of \cite{wd05}. The dark
matter halo had a virial mass $M_{\rm H}=7.7 \times 10^{11} \rm{M}_{\odot}$ and
concentration $c=27$. The disk had a mass $M_{\rm D}=3.4 \times 10^{10}\rm{M}_{\odot}$,
the scale-length $R_{\rm D}=2.82$ kpc and thickness $z_{\rm
D}=0.44$ kpc. The initial conditions fulfilled the Toomre's stability
criterion ($Q>2$).

The Virgo cluster was approximated as an NFW dark matter halo
with parameters estimated by \cite{virgo1} and \cite{virgo2}, namely the virial
mass $M_{\rm C}=5.4 \times 10^{14}$ M$_{\odot}$ and concentration $c=3.8$.
The $N$-body realizations for both, the galaxy and the cluster, were generated
via procedures described in \cite{wd05} and consisted of $10^6$
particles per component.

The galaxy was placed at an apocenter of a typical eccentric orbit in the Virgo cluster with an
apo- to pericenter distance ratio of $r_{\rm apo}/r_{\rm
peri}=1500/300$ kpc. The evolution was followed for $10$ Gyr with the GADGET-2 $N$-body
code (\citealt{gadget2}).

\section{Evolution and structure of spiral arms}

\begin{figure}
\includegraphics[width=1.2\textwidth]{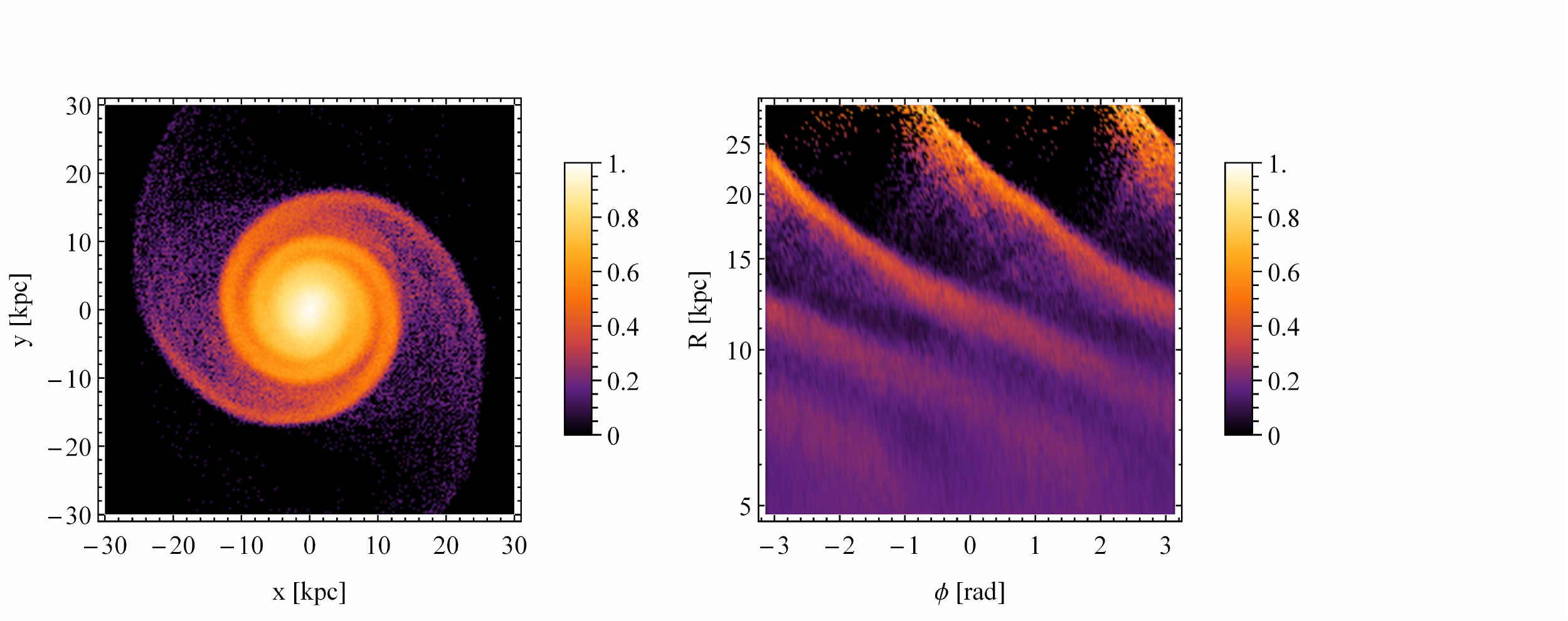}
\caption{{\it Left:} Face-on view of the surface density distribution of stars in the disk
at 2.75 Gyr. The color bar labels were normalized to $0.32 \log[1+\Sigma/(3.76 \times 10^5
\rm{M}_{\odot}/\rm{kpc}^2)]$. {\it Right:} Perturbed surface density at the same time in the
$\phi-\ln R$ plane. The color bar labels were normalized to $0.53 \log[2+(\Sigma-\Sigma_0)/\Sigma_0].$ }
\label{fig1}
\end{figure}

Grand design, two-armed structures form after the first pericenter passage of the galaxy on its orbit in
the Virgo cluster. The left panel of Fig.~\ref{fig1} shows the face-on view of the surface density of the disk at 2.75
Gyr, i.e. 0.85 Gyr after the first pericenter. The right panel of the Figure shows the perturbed density
$(\Sigma-\Sigma_0)/\Sigma_0$ (where $\Sigma$ is the surface density at a given time and $\Sigma_0$ is the initial
value) at the same time in the $\phi-\ln R$ plane. The right-panel plot demonstrates
that the spiral arms are approximately logarithmic.

\begin{figure}
\begin{center}
\includegraphics[width=0.9\textwidth]{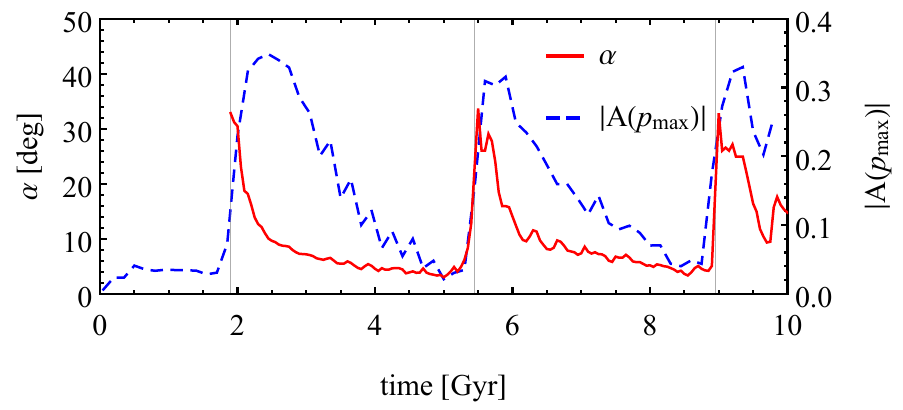}
\caption{The evolution of the pitch angle $\alpha$ and the arm strength $|A(p_{\rm max})|$ measured in the ring of 9
kpc $\leq$ $R$ $\leq$ 15 kpc. Gray vertical lines indicate pericenter passages.}
\end{center}
\label{fig2}
\end{figure}

The spiral structure formed shortly after the pericenter passage
is not stable, but the arms wind up with time. After about 2 Gyr they
dissolve to be triggered again during the next pericenter passage. This recurrent,
transient behaviour is confirmed by the measurements of the pitch angle and the arm strength.

We expanded the surface distribution of stars in the ring 9 kpc $\leq$ $R$ $\leq$ 15 kpc
in logarithmic spirals according to the formula
$A(p) = (1/N_{\rm s}) \Sigma_{j} \exp [i (2 \phi_j +p \ln R_j)]$ where $N_{\rm s}$ is the number of stars and
$(\phi_j, R_j)$ are the coordinates of the $j$th star
\citep[see e.g.][]{sa86}. We then find $p_{\rm max}$ that maximizes $|A(p)|$ and the pitch angle $\alpha$ from
$\tan \alpha=2/p_{\rm max}$.
The measurements are shown
as a function of time with the red line in Fig.~\ref{fig2}. Just after the pericenters, when the
arms form, $\alpha$
is as high as $\sim35^{\circ}$ and then it exponentially decreases to $\sim5^{\circ}$. This
decrease of pitch angle corresponds to the arms winding up during the evolution.
\enlargethispage{\baselineskip}

The Figure also shows the arm strength $|A(p_{\rm max})|$ as a function of time with the dashed blue line.
The arm strength clearly rises during and just after pericenters and then decreases. This means that arms are
dissolving but they do not entirely vanish before the next pericenter.

\section{Summary}

We performed $N$-body simulations of a Milky Way-like galaxy orbiting in a
Virgo-like cluster. We found that tidal forces from the cluster induce the formation of grand-design
spiral arms at pericenter passages. The spiral arms wind up and weaken with time until the next pericenter when
they are recreated.

\acknowledgements{This work was partially supported by the Polish National Science Centre
under grant 2013/10/A/ST9/00023.}

\bibliographystyle{ptapap}
\bibliography{draft_ewa}

\end{document}